\documentclass[12pt]{iopart}
\usepackage{epsfig,psfig,cite,amsfonts}

\newcommand{\gsim}{\lower.7ex\hbox{$\;\stackrel{\textstyle>}
{\sim}\;$}}
\catcode`@=11
\def\citer{\@ifnextchar [{\@tempswatrue\@citexr}{\@tempswafalse\@citexr[]}}
 
\def\@citexr[#1]#2{\if@filesw\immediate\write\@auxout{\string\citation{#2}}\fi
  \def\@citea{}\@cite{\@for\@citeb:=#2\do
    {\@citea\def\@citea{--\penalty\@m}\@ifundefined
       {b@\@citeb}{{\bf ?}\@warning
       {Citation `\@citeb' on page \thepage \space undefined}}%
\hbox{\csname b@\@citeb\endcsname}}}{#1}}
\catcode`@=12

\def\refeq#1{\mbox{eq.~(\ref{#1})}}

\def\reffi#1{\mbox{Fig.~\ref{#1}}}

\def\citere#1{\mbox{Ref.~\cite{#1}}}

\newcommand{\MstL}{M_{\tilde{t}_L}}
\newcommand{\MstR}{M_{\tilde{t}_R}}

\newcommand{\Xt}{X_t}

\newcommand{\msusy}{M_{\mathrm{SUSY}}}

\newcommand{\cp}{{\cal CP}}

\newcommand{\twol}{two-loop}
\newcommand{\onel}{one-loop}

\newcommand{\MA}{M_A}
\newcommand{\mh}{m_h}
\newcommand{\mhmax}{m_h^{\rm max}}

\newcommand{\mt}{m_{t}}

\newcommand{\mgl}{m_{\tilde{g}}}

\newcommand{\tsf}{\theta\kern-.20em_{\tilde{f}}}
\newcommand{\tsfp}{\theta\kern-.20em_{\tilde{f}\prime}}
\newcommand{\tsq}{\theta\kern-.15em_{\tilde{q}}}

\newcommand{\VL}{\left( \begin{array}{c}}
\newcommand{\VR}{\end{array} \right)}
\newcommand{\ML}{\left( \begin{array}{cc}}
\newcommand{\MLd}{\left( \begin{array}{ccc}}
\newcommand{\MLv}{\left( \begin{array}{cccc}}
\newcommand{\MR}{\end{array} \right)}

\newcommand{\tb}{\tan \beta}

\newcommand{\tev}{\,\, \mathrm{TeV}}
\newcommand{\gev}{\,\, \mathrm{GeV}}

\def\Gcs{\mathrm{GeV}}

\newcommand{\BC}{\begin{center}}
\newcommand{\EC}{\end{center}}
\newcommand{\BE}{\begin{equation}}
\newcommand{\EE}{\end{equation}}
\newcommand{\BEA}{\begin{eqnarray}}
\newcommand{\BEAnn}{\begin{eqnarray*}}
\newcommand{\EEA}{\end{eqnarray}}
\newcommand{\EEAnn}{\end{eqnarray*}}
\newcommand{\non}{\nonumber}
\newcommand{\id}{{\rm 1\kern-.12em
\rule{0.3pt}{1.5ex}\raisebox{0.0ex}{\rule{0.1em}{0.3pt}}}}
\newcommand{\lesim}
{\;\raisebox{-.3em}{$\stackrel{\displaystyle <}{\sim}$}\;}
\def\De{\Delta}

\begin{document}

\thispagestyle{empty}
\setcounter{page}{0}
\def\thefootnote{\fnsymbol{footnote}}

\begin{flushright}
CERN--TH/99--368\\
DESY 99--156\\
RAL--TR--1999--079 \\
GLAS-PPE/1999--20 \\
hep-ph/9912249 \\
\end{flushright}

\vspace{0.5cm}

\begin{center}

{\large\sc {\bf Upper limit on 
$\mh$ in the MSSM and M-SUGRA }}

\vspace*{0.4cm} 

{\large\sc {\bf vs.\ prospective reach of LEP}}%
\footnote[7]{
\footnoterule
Contribution to the ``UK Phenomenology Workshop on 
          Collider Physics'',\\ \mbox{}~~~Durham, UK, September 1999.}

\vspace{0.7cm}

{\sc A.~Dedes$^{\,1}$%
\footnote[2]{
email: dedesa@hephp4.pp.rl.ac.uk
}%
, S.~Heinemeyer$^{\,2}$%
\footnote[8]{
email: Sven.Heinemeyer@desy.de
}%
, P.~Teixeira-Dias$^{\,3}$%
\footnote[6]{
email: PedroTD@physics.gla.ac.uk
}%
~and G.~Weiglein$^{\,4}$%
\footnote[1]{
email: Georg.Weiglein@cern.ch
}%
}

\vspace*{1cm}

$^1$ Rutherford Appleton Laboratory, Chilton, Didcot, Oxon OX11 0QX, UK

\vspace*{0.2cm}

$^2$  DESY Theorie, Notkestr. 85, 22603 Hamburg, Germany

\vspace*{0.2cm}

$^3$ Dept. of Physics and Astronomy, University of Glasgow, Glasgow G12 8QQ, UK

\vspace*{0.2cm}

$^4$ CERN, TH Division, CH-1211 Geneva 23, Switzerland
\end{center}

\vspace*{0.7cm}

\BC
{\bf Abstract}
\EC
The upper limit on the lightest $\cp$-even Higgs boson mass, $\mh$,
is analyzed within the MSSM as a function of $\tb$ for fixed $\mt$ and
$\msusy$. The impact of recent diagrammatic \twol\ results on this limit
is investigated. We compare the MSSM theoretical upper bound on $\mh$
with the lower bound obtained from experimental searches at LEP. We
estimate that with the LEP data taken until the end of 1999, the region
$\mh<108.2~\Gcs$ can be excluded at the 95\% confidence level. This
corresponds to an excluded region $0.6\lesim\tb\lesim 1.9$ within the MSSM
for $\mt = 174.3 \gev$ and $\msusy \leq 1$~TeV. 
The final exclusion sensitivity after the end of
LEP, in the year 2000, is also briefly discussed. Finally, we determine
the upper limit on $\mh$ within the Minimal Supergravity (M-SUGRA)
scenario up to the \twol\ level, consistent with radiative
electroweak symmetry breaking. We find an upper bound of $\mh \approx
127 \gev$ for $\mt = 174.3 \gev$ in this scenario, which is slightly
below the bound in the unconstrained MSSM.
%\end{abstract}

\def\thefootnote{\arabic{footnote}}
\setcounter{footnote}{0}

\newpage

%%%%%%%%%%%%%%%%%%%%%%%%%%%%%%%%%%%%%%%%%%%%%%%%%%%%%%%%%%%%%%%%%%%%
%%%%%%%%%%%%%%%%%%%%%%%%%%%%%%%%%%%%%%%%%%%%%%%%%%%%%%%%%%%%%%%%%%%%

\title{Upper limit on 
$\mh$ in the MSSM and M-SUGRA vs.\ prospective reach of LEP}

\author{A. Dedes\dag, S. Heinemeyer\P, P. Teixeira-Dias* and~G. Weiglein\S}

\address{
\dag\ Rutherford Appleton Laboratory, Chilton, Didcot, Oxon OX11 0QX, UK \\
\P\ DESY Theorie, Notkestr. 85, 22603 Hamburg, Germany \\
$*$ Dept. of Physics and Astronomy, University of Glasgow, Glasgow G12 8QQ, UK\\
\S\ CERN, TH Division, CH-1211 Geneva 23, Switzerland}

\begin{abstract}
The upper limit on the lightest $\cp$-even Higgs boson mass, $\mh$,
is analyzed within the MSSM as a function of $\tb$ for fixed $\mt$ and
$\msusy$. The impact of recent diagrammatic \twol\ results on this limit
is investigated. We compare the MSSM theoretical upper bound on $\mh$
with the lower bound obtained from experimental searches at LEP. We
estimate that with the LEP data taken until the end of 1999, the region
$\mh<108.2~\Gcs$ can be excluded at the 95\% confidence level. This
corresponds to an excluded region $0.6\lesim\tb\lesim 1.9$ within the MSSM
for $\mt = 174.3 \gev$ and $\msusy \leq 1$~TeV. 
The final exclusion sensitivity after the end of
LEP, in the year 2000, is also briefly discussed. Finally, we determine
the upper limit on $\mh$ within the Minimal Supergravity (M-SUGRA)
scenario up to the \twol\ level, consistent with radiative
electroweak symmetry breaking. We find an upper bound of $\mh \approx
127 \gev$ for $\mt = 174.3 \gev$ in this scenario, which is slightly
below the bound in the unconstrained MSSM.
\end{abstract}

%\pacs{...}

%%%%%%%%%%%%%%%%%%%%%%%%%%%%%%%%%%%%%%%%%%%%%%%%%%%%%%%%%%%%%%%%%%%%%%%%%
%%%%%%%%%%%%%%%%%%%%%%%%%%%%%%%%%%%%%%%%%%%%%%%%%%%%%%%%%%%%%%%%%%%%%%%%%

\section{Introduction}

Within the MSSM the masses of the $\cp$-even neutral Higgs bosons are
calculable in terms of the other MSSM parameters. The mass of the
lightest Higgs boson, $\mh$, has been of particular interest, as
it is bounded to be smaller than the $Z$~boson mass at the tree level. 
The \onel\ results~\cite{mhiggs1l,mhiggsf1l,mhiggsf1ldab,pierce} 
for $\mh$ have been supplemented in the
last years with the leading \twol\ corrections, performed in the
renormalization group (RG)
approach~\cite{mhiggsRG1,mhiggsRG2}, in the effective
potential approach~\cite{mhiggsEP} and most recently in
the Feynman-diagrammatic (FD)
approach~\cite{mhiggsletter,mhiggslong}. 
The \twol\ corrections have turned out to be sizeable. They can
change the \onel\ results by up to 20\%.

Experimental searches at LEP now exclude a light MSSM Higgs boson
with a mass below 
$\sim$90~GeV~\cite{lepc-aleph,lepc-delphi,lepc-l3,lepc-opal}.
In the low $\tb$ region, in which the limit is the same as for the 
Standard Model Higgs boson, a mass limit of even $\mh \gsim 106 \gev$ has
been obtained~\cite{lepc-aleph,lepc-delphi,lepc-l3,lepc-opal}.
Combining this experimental bound with the theoretical upper limit 
on $\mh$ as a function of $\tb$ within the MSSM, it is possible to derive
constraints on $\tb$. In this paper we investigate, for which MSSM
parameters the maximal $\mh$ values are obtained and discuss in this
context the impact of the new FD two-loop result. Resulting constraints
on $\tb$ are analyzed on the basis of the present LEP data and of the
prospective final exclusion limit of LEP.

The Minimal Supergravity (M-SUGRA) scenario provides a relatively
simple and constrained version of the MSSM. 
In this paper we explore, how the maximum possible values for $\mh$ change
compared to the general MSSM, if one restricts to the M-SUGRA framework. 
As an additional constraint we impose that the condition of radiative 
electroweak symmetry breaking (REWSB)~\cite{REWSB} should be fulfilled. 

%%%%%%%%%%%%%%%%%%%%%%%%%%%%%%%%%%%%%%%%%%%%%%%%%%%%%%%%%%%%%%%%%%%%%%%%%
%%%%%%%%%%%%%%%%%%%%%%%%%%%%%%%%%%%%%%%%%%%%%%%%%%%%%%%%%%%%%%%%%%%%%%%%%

\section{The upper bound on $\mh$ in the MSSM} 
\label{section:mssm}

The most important radiative corrections to $\mh$ arise from the top and
scalar top sector of the MSSM, with the input parameters $\mt$, $\msusy$
and $\Xt$. Here we assume the soft SUSY breaking parameters in the
diagonal entries of the scalar top mixing matrix to be equal for simplicity,
$\msusy = \MstL = \MstR$. This has been shown to yield upper values for 
$\mh$ which comprise also the case where $\MstL \neq \MstR$,
if $\msusy$ is identified with the heavier one of $\MstL$, 
$\MstR$~\cite{mhiggslong}. For the off-diagonal entry of the mixing
matrix we use the convention 
\BE
\label{eq:xt}
\mt \Xt = \mt (A_t - \mu \cot\beta) .
\EE
Note that the sign convention used for $\mu$ here is the opposite of the
one used in \citere{sakis}.

Since the predicted value of $\mh$ depends
sensitively on the precise numerical value of $\mt$, it has become
customary to discuss the constraints on $\tb$ within a so-called
``benchmark'' scenario (see \citere{lephiggs183} and references therein), 
in which $\mt$ is kept fixed at the value $\mt = 175 \gev$ and in which 
furthermore a large value of $\msusy$ is chosen,
$\msusy = 1 \tev$, giving rise to large values of $\mh(\tb)$. 
In \citere{tbexcl} it has recently been analyzed how the values chosen
for the other SUSY parameters in the benchmark scenario should be modified
in order to obtain the maximal values of $\mh(\tb)$ for given $\mt$ and
$\msusy$. The corresponding scenario ($\mhmax$ scenario) is defined 
as~\cite{tbexcl,bench}
\BEA
&& \mt = \mt^{\mathrm{exp}}~  ( = 174.3 \gev ), \quad \msusy = 1 \tev \non \\
&& \mu = -200 \gev, \; M_2 = 200 \gev, \; \MA = 1 \tev, 
   \; \mgl = 0.8 \, \msusy ({\rm FD}) \non \\
&& \Xt = 2\, \msusy ({\rm FD})\;\; {\rm or}\;\; 
   \Xt = \sqrt{2}\, \msusy ({\rm RG}) ,
\label{benchmarkdef}
\EEA
where the parameters are chosen such that the chargino masses are beyond
the reach of LEP2 and that the lightest $\cp$-even Higgs boson does not
dominantly decay invisibly into neutralinos. In \refeq{benchmarkdef}
$\mu$ is the Higgs mixing parameter, $M_2$ denotes the
soft SUSY breaking parameter in the gaugino sector, and $\MA$ is the
$\cp$-odd Higgs boson mass. The gluino mass, $\mgl$, can only be
specified as a free parameter 
in the FD result (program {\tt FeynHiggs}~\cite{feynhiggs}).
The effect of varying $\mgl$ on $\mh$ is up to $\pm 2
\gev$~\cite{mhiggslong}. Within the RG result 
(program {\tt subhpole}~\cite{mhiggsRG1}) $\mgl$ is fixed to $\mgl = \msusy$. 
Compared to the maximal values for $\mh$ (obtained for $\mgl \approx
0.8\,\msusy$) this leads to a reduction of the Higgs boson mass by up to
$0.5 \gev$. Different values of $\Xt$ are specified in
\refeq{benchmarkdef} for the results of the FD and the RG calculation,
since within the two approaches the maximal values for $\mh$ are
obtained for different values of $\Xt$. This fact is partly due to the
different renormalization schemes used in the two approaches~\cite{bse}.

The maximal values for $\mh$ as a function of $\tb$ within the
$\mhmax$ scenario are higher by about 5~GeV than in the
previous benchmark scenario. The constraints on $\tb$ derived within the
$\mhmax$ scenario are thus more conservative than the ones based on the
previous scenario.

The investigation of the constraints on $\tb$ that can be obtained from the 
experimental search limits on $\mh$ has so far been based on the 
results for $\mh$ obtained within the RG approach~\cite{mhiggsRG1}.
The recently obtained FD~\cite{mhiggsletter,mhiggslong} result differs 
from the RG result by a more complete treatment of the one-loop
contributions~\cite{mhiggsf1ldab} and in particular by genuine
non-logarithmic two-loop terms that go beyond the leading logarithmic 
two-loop contributions contained in the RG result~\cite{bse,mhiggslle}. 
Comparing the FD result (program {\tt FeynHiggs}) with
the RG result (program {\tt subhpole}) we find that the
maximal value for $\mh$ as a function of $\tb$ within the FD result is
higher by up to 4~GeV.

\begin{figure}[ht!]
%\vspace{1em}
\begin{center}
\mbox{
\psfig{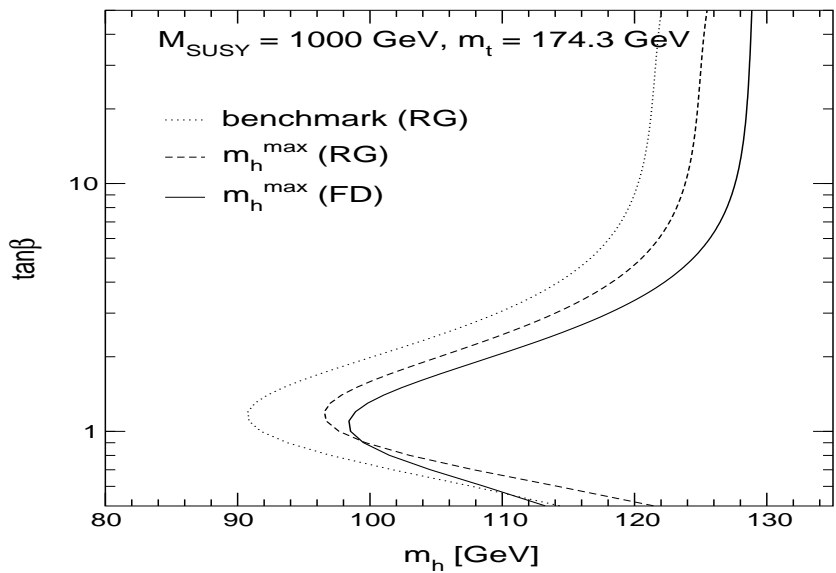}}
\end{center}
\caption[]{
The upper bound on $\mh$ is shown as a function of $\tb$ for given $\mt$
and $\msusy$.
The dashed curve displays the previous benchmark scenario. 
The dotted curve shows the RG result for the $\mhmax$ scenario, 
while the solid curve represents the FD result for the $\mhmax$ scenario.
}
\label{fig:rgv}
\end{figure}

In \reffi{fig:rgv} we show both the effect of modifying the previous
benchmark scenario to the $\mhmax$ scenario and the impact of the new FD
two-loop result on the prediction for $\mh$. The maximal value for the 
Higgs boson mass is plotted as a function of $\tb$ for $\mt = 174.3$~GeV
and $\msusy = 1$~TeV. The dashed curve displays the benchmark scenario,
used up to now by the LEP collaborations~\cite{lephiggs183}. The dotted
curve shows the $\mhmax$ scenario. Both curves are based on the RG
result (program {\tt subhpole}). The solid curve corresponds to the FD
result (program {\tt FeynHiggs}) in the $\mhmax$ scenario. The increase 
in the maximal value for $\mh$ by about $4$~GeV from the new FD result 
and by further 5~GeV if the benchmark scenario is replaced by the 
$\mhmax$ scenario has a
significant effect on exclusion limits for $\tb$ derived from the
Higgs boson search. Combining both effects, which of course have a very
different origin, the maximal Higgs boson masses are increased by almost
$10 \gev$ compared to the previous benchmark scenario.

%%%%%%%%%%%%%%%%%%%%%%%%%%%%%%%%%%%%%%%%%%%%%%%%%%%%%%%%%%%%%%
%%%%%%%%%%%%%%%%%%%%%%%%%%%%%%%%%%%%%%%%%%%%%%%%%%%%%%%%%%%%%%

{}From the FD result we find the upper bound of $\mh \lesim 129$~GeV in
the region of large $\tb$ within the MSSM for $\mt = 174.3 \gev$ and
$\msusy = 1$~TeV. Higher values for $\mh$ are
obtained if the experimental uncertainty in $\mt$ of currently 
$\De\mt = 5.1 \gev$ is taken into account and higher values are allowed 
for the top quark mass. As a rule of thumb, increasing $\mt$ by 1~GeV
roughly translates into an upward shift of $\mh$ of 1~GeV. An
increase of $\msusy$ from 1~TeV to 2~TeV enhances $\mh$ by about 2~GeV
in the large $\tb$ region. As an extreme case, choosing $\mt =
184.5$~GeV, i.e.\ two standard deviations above the current experimental
central value, and using $\msusy = 2$~TeV leads to an upper bound on 
$\mh$ of $\mh \lesim 141$~GeV within the MSSM.

%%%%%%%%%%%%%%%%%%%%%%%%%%%%%%%%%%%%%%%%%%%%%%%%%%%%%%%%%%%%%%%%%%%%%%%%%
%%%%%%%%%%%%%%%%%%%%%%%%%%%%%%%%%%%%%%%%%%%%%%%%%%%%%%%%%%%%%%%%%%%%%%%%%

%%%%%%%%%%%%%%%%%%%%%%%%%%%%%%%%%%%%%%%%%%%%%%%%%%%%%%%%%%%%%%%%%%%%%%%%%
%%%%%%%%%%%%%%%%%%%%%%%%%%%%%%%%%%%%%%%%%%%%%%%%%%%%%%%%%%%%%%%%%%%%%%%%%

\section{The prospective upper $\mh$ reach of LEP}

The four LEP experiments are very actively searching for the Higgs
boson.  Results presented recently by the LEP collaborations revealed
no evidence of a SM Higgs boson signal in the data collected in 1999
at centre-of-mass energies of approximately 192, 196, 200 and 202
GeV\cite{lepc-aleph,lepc-delphi,lepc-l3,lepc-opal}. From the negative results of their searches ALEPH, DELPHI and L3 
have therefore individually excluded a SM Higgs boson lighter than
$\sim$101--106 $\Gcs$ (at the 95\% confidence 
level)~\cite{lepc-aleph,lepc-delphi,lepc-l3}.

Here we will present the expected exclusion reach of LEP assuming all
the data taken by the four experiments in 1999 is combined. The
ultimate exclusion reach of LEP -- assuming no signal were found in
the data to be collected in the year 2000 -- will also be estimated
for several hypothetical scenarios of luminosity and centre-of-mass
energy. These results are then confronted with the theoretical MSSM
upper limit on $\mh(\tb)$ presented in Section \ref{section:mssm},
in order to establish to what extent the LEP data can probe the
low $\tb$ region.  We recall that models in which b-$\tau$ Yukawa
coupling unification at the GUT scale is imposed favor low $\tb$
values, $\tb \approx 2$, which can %possibly be ruled out 
severely be constrained experimentally by searches at LEP. Alternatively, 
such models can favor $\tb \approx 40$, a region which however can only 
be partly covered at LEP.

 All experimental exclusion limits quoted in this section are
 implicitly meant at the 95\% confidence level (CL).

It has been proposed \cite{safari} that the LEP-combined expected 95\% CL
lower bound on $\mh$, $\mh^{95}$, for a data set consisting of data
accumulated at given centre-of-mass energies can be estimated by
solving the equation
\BE  
 n(\mh^{95}) = (\sigma_0 \mathcal{L}_{eq})^{\alpha}, \label{eqn:predictor}
\EE
 where $n(\mh^{95})$ is the number of signal events produced at the
 95\% CL limit. The equivalent luminosity, $\mathcal{L}_{eq}$, is
 the luminosity that one would have to accumulate at the highest
 centre-of-mass energy in the data set in order to have the same
 sensitivity as in the real data set, where the data is split between
 several different $\sqrt{s}$ values. For a SM Higgs boson signal, the
 parameters $\sigma_{0}$ and $\alpha$ are $\sim$38 pb and $\sim$0.4,
 respectively \cite{safari}. (These parameter values are obtained from
 a fit to the actual LEP-combined expected limits from
 $\sqrt{s}=161$ GeV up to $\sqrt{s}=188.6$ GeV
 \cite{lephiggs172,lephiggs183,lephiggs189}.) The predicted $\mh$
 limits obtained with this method are expected to approximate the more
 accurate combinations done by the LEP Higgs Working Group, with
 an uncertainty of the order of $\pm$ 0.3 $\Gcs$.

\begin{figure}[htbp]
%{\epsfig{file=predictor.eps,width=15.0cm}}
\begin{tabular}{rlr}
{\epsfig{file=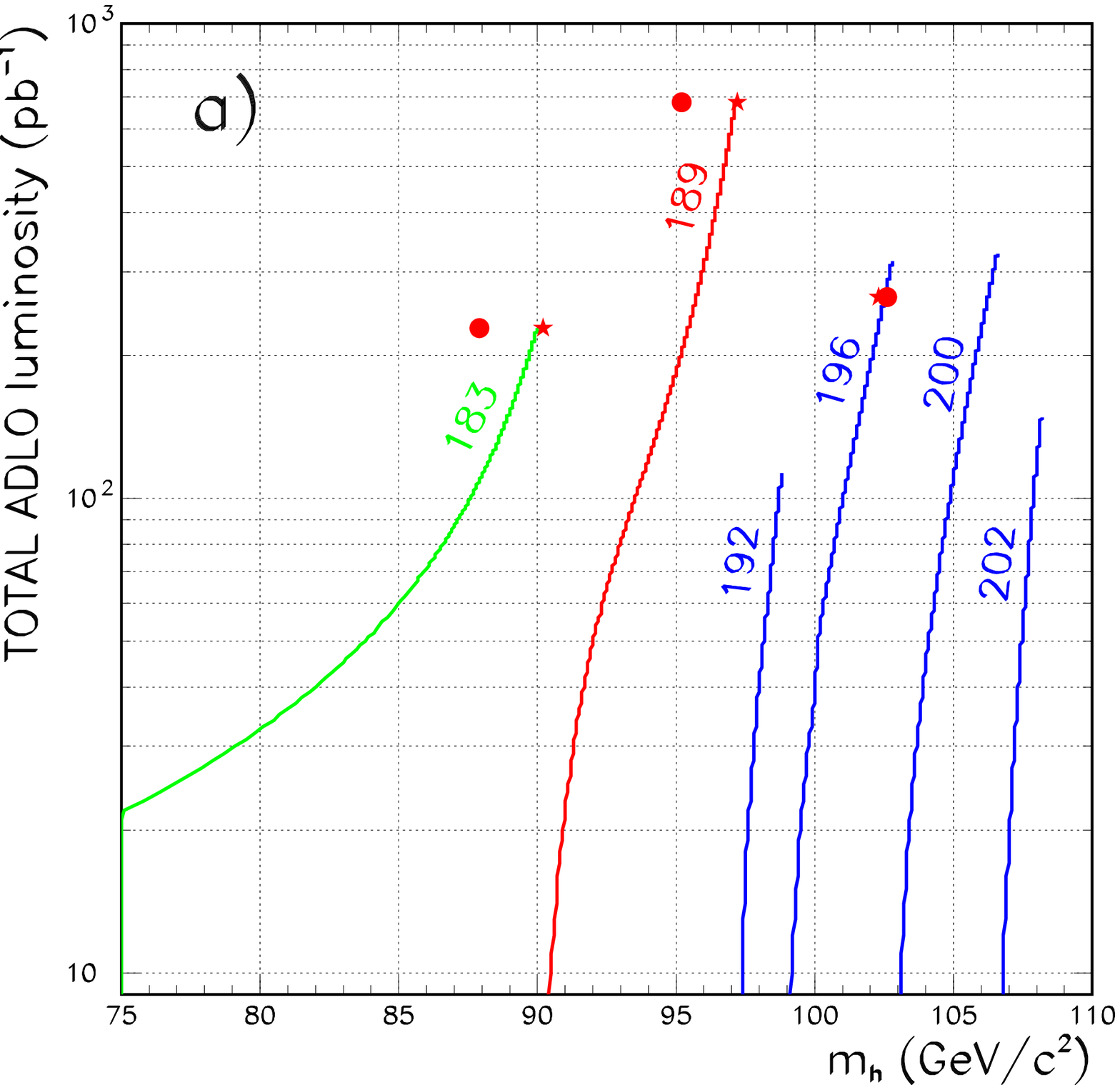,width=7.9cm}}
{\epsfig{file=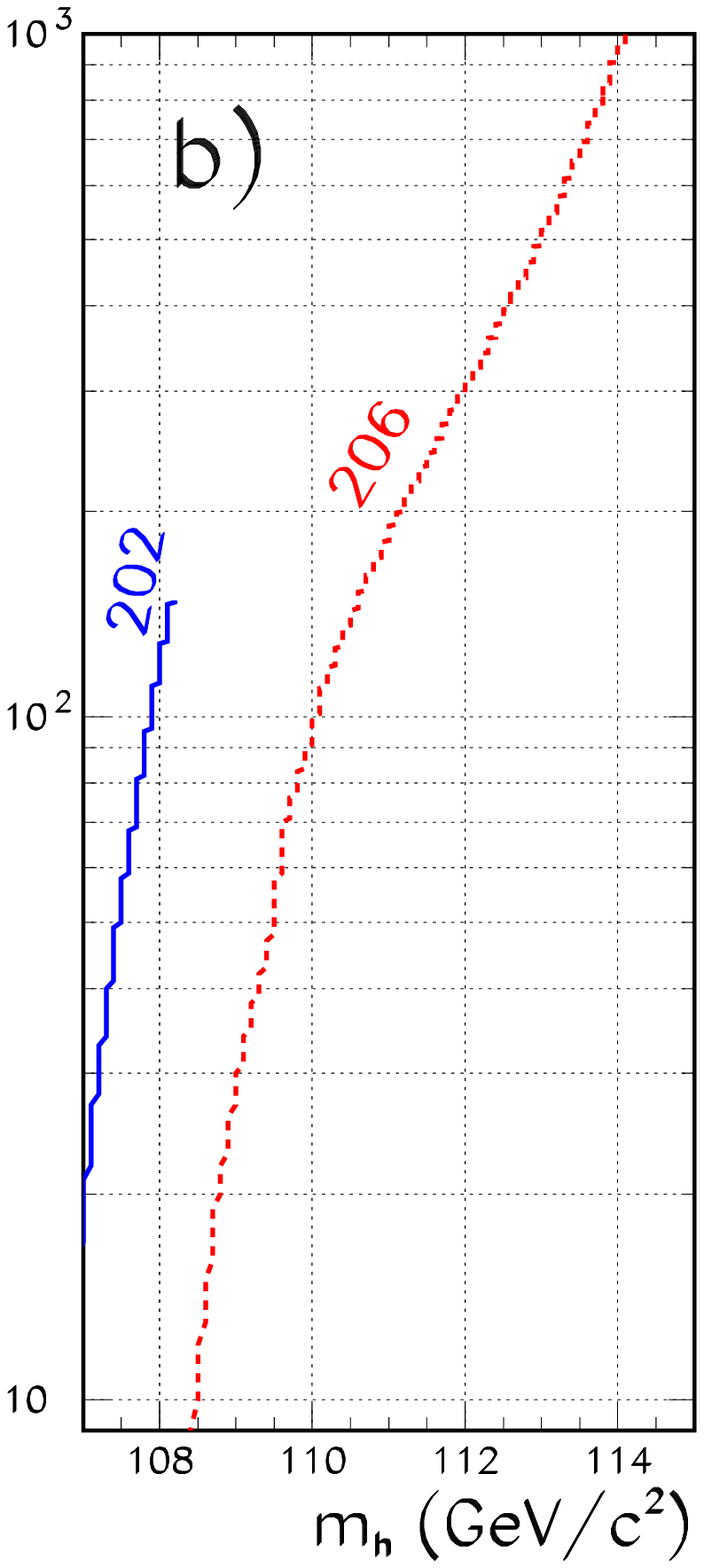,width=3.7cm}}
{\epsfig{file=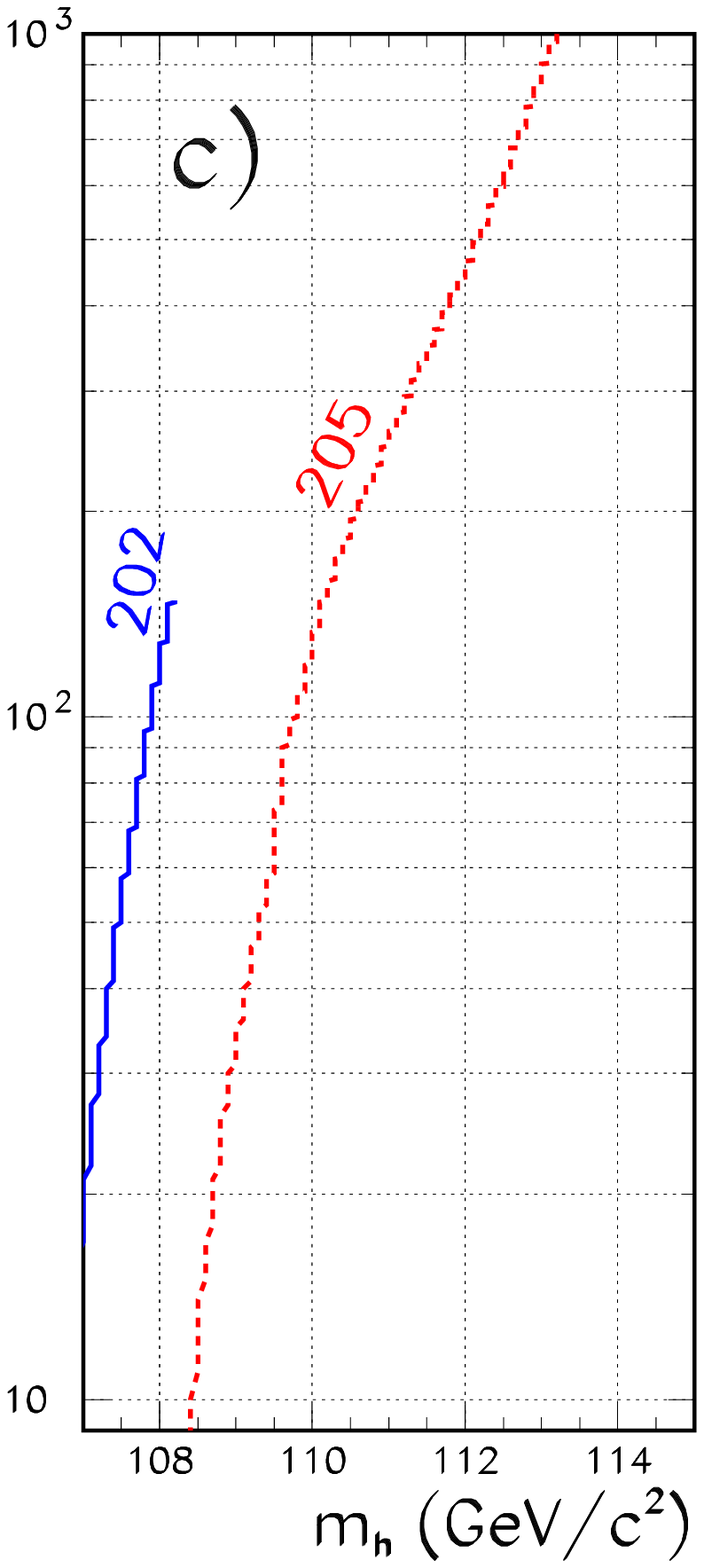,width=3.7cm}}
\end{tabular}
\caption{
{\footnotesize Predictions of the expected combined
ALEPH+DELPHI+L3+OPAL 95\% CL $\mh$ exclusion; a) obtained from the
data taken until the end of 1999 (solid lines). For comparison the
expected (stars) and observed (dots) combined LEP limits obtained from
actual data combinations\cite{lephiggs183,lephiggs189,lephiggs196} are
also shown. The effect of adding to this data set new data at
b) $\sqrt{s}=$206 GeV or c) 205 GeV is indicated by the dashed line.
\label{fig:predictor}}
}
\end{figure}

 Solving \refeq{eqn:predictor} for the existing LEP data with
 183 GeV $\lesim \sqrt{s} \lesim 202 $ GeV (Table~\ref{tab:lepdata})
 results in a predicted combined exclusion of $\mh < 108.2~\Gcs$ for
 the SM Higgs boson (see Figure \ref{fig:predictor}a).

% --- table
\begin{table}[htbp]
\caption{
{\footnotesize Summary of the total LEP data luminosity accumulated
since 1997. The luminosities for the data taken in 1999 ($\sqrt{s}\ge$
191.6 GeV) are the (still preliminary) values
quoted by the four LEP experiments at the LEPC open session
\cite{lepc-aleph,lepc-delphi,lepc-l3,lepc-opal}.}
\label{tab:lepdata} 
}
\begin{center}
\begin{tabular}{|c|cccccc|}
\hline
 $\sqrt{s}$ (GeV)          & 182.7 & 188.6 & 191.6 & 195.5 & 199.5 & 201.6 \\
\hline
 $\mathcal{L}$ (pb$^{-1}$) & 220.0  & 682.7 & 113.9 & 316.4 & 327.8 & 148.1 \\
\hline
\end{tabular}
\end{center}
\end{table}

 Based on the current LEP operational experience, it is believed that
 in the year 2000 stable running is possible up to $\sqrt{s}=206$
 GeV\cite{lepc-lep}. Figure \ref{fig:predictor}b demonstrates the
 impact of additional data collected at $\sqrt{s}=206$ GeV
 on the exclusion. For instance, if no evidence of a signal were
 found in the data, collecting 500 (1000) pb$^{-1}$ at this
 centre-of-mass energy would increase the $\mh$ limit to 113.0 (114.1)
 $\Gcs$. Figure \ref{fig:predictor}c shows the degradation in the
 sensitivity to a Higgs boson signal if the data in the year 2000 were
 accumulated at $\sqrt{s}=205$ GeV instead: in this case the
 luminosity required to exclude up to $\mh=113~\Gcs$ would be
 840~pb$^{-1}$.

% --- figure

% --- table
\begin{table}[htbp]
\caption{
{\footnotesize Predictions of the sensitivity of the four LEP
experiments combined, for several hypothetical data sets. The table
shows the expected excluded SM Higgs boson mass ($\mh^{95}$, in GeV) as well
as the corresponding excluded $\tb$ region in the $\mhmax$ benchmark
scenario (with $\mt = 174.3$~GeV, $\msusy = 1$~TeV),
when new data at the indicated $\sqrt{s}$ is combined with the
existing data set (Table
\ref{tab:lepdata}). The luminosities indicated are for the 4 LEP experiments 
combined. The results shown are valid only if no signal were found in
the data. (Note that, as it is not foreseen at the moment that it will
be possible to run LEP at
$\sqrt{s}>206$ GeV, scenario 8 is probably unrealistic.) }
\label{tab:lepreach} 
}
\begin{center}
\begin{tabular}{|c||cccc||cc|}
\hline
 $\sqrt{s}$ (GeV)             & 204. & 205. & 206. & 208. & $\mh^{95}$ & $\tb^{95}$ \\
\hline \hline
 1) $\mathcal{L}$ (pb$^{-1}$) &  -   &  -   & 100. &  -   & 110.0 & 0.6 -- 2.1 \\
\hline
 2) $\mathcal{L}$ (pb$^{-1}$) &  -   &  -   & 500. &  -   & 113.0 & 0.5 -- 2.4 \\
\hline
 3) $\mathcal{L}$ (pb$^{-1}$) &  -   &  -   & 1000.&  -   & 114.1 & 0.5 -- 2.5 \\
\hline
 4) $\mathcal{L}$ (pb$^{-1}$) &  -   & 120. &  -   &  -   & 110.0 & 0.6 -- 2.1 \\
\hline
 5) $\mathcal{L}$ (pb$^{-1}$) &  -   & 840. &  -   &  -   & 113.0 & 0.5 -- 2.4 \\
\hline
 6) $\mathcal{L}$ (pb$^{-1}$) & 100. & 100. & 400. &  -   & 113.1 & 0.5 -- 2.4 \\
\hline
 7) $\mathcal{L}$ (pb$^{-1}$) & 150. & 300. & 300. &  -   & 113.3 & 0.5 -- 2.4 \\
\hline
 8) $\mathcal{L}$ (pb$^{-1}$) & 150. & 300. & 300. & 280. & 115.0 & 0.5 -- 2.6 \\
\hline
\end{tabular}
\end{center}
\end{table}

 In Table \ref{tab:lepreach} the expected SM Higgs boson limit is
 shown for several possible LEP running scenarios in the year
 2000. Taking into account that the {\sl experimental} MSSM $\mh$
 exclusion in the range $0.5 \lesim \tb \lesim 3$ is {\sl (i)}
 essentially independent of $\tb$ and {\sl (ii)} equal in value to the
 SM $\mh$ exclusion (see e.g. \cite{lephiggs189,lephiggs196}), 
 $\mh^{95}$ can be
 converted into an excluded $\tb$ range in the $\mhmax$ benchmark
 scenario described in Section \ref{section:mssm}.
 This is done by intersecting
 the experimental exclusion and the solid curve in Figure
 \ref{fig:rgv}. Using the LEP data taken until the end of 1999 (for
 which $\mh^{95}=108.2~\Gcs$) one can already expect to exclude 
 $0.6 \lesim \tb \lesim 1.9$ 
 within the MSSM for $\mt = 174.3$~GeV and $\msusy = 1$~TeV. 
 Note that in determining the excluded $\tb$ regions in Table
 \ref{tab:lepreach} the theoretical uncertainty from unknown
 higher-order corrections has been neglected.
 As can be seen from Table \ref{tab:lepreach}, several
 plausible scenarios for adding new data at higher energies can extend
 the exclusion to $\mh\lesim 113~\Gcs$ ($0.5 \lesim \tb \lesim 2.4$).

%%%%%%%%%%%%%%%%%%%%%%%%%%%%%%%%%%%%%%%%%%%%%%%%%%%%%%%%%%%%%%%%%%%%%%%%%
%%%%%%%%%%%%%%%%%%%%%%%%%%%%%%%%%%%%%%%%%%%%%%%%%%%%%%%%%%%%%%%%%%%%%%%%%

\section{The upper limit on $\mh$ in the M-SUGRA scenario}

The M-SUGRA scenario is described by four independent parameters and a
sign,
namely the common squark mass $M_0$, the common gaugino mass $M_{1/2}$, 
the common trilinear coupling $A_0$, $\tb$ and the sign of $\mu$.
The universal parameters are fixed at the GUT scale, where we assumed
unification of the gauge couplings. Then they are run down
to the electroweak scale with the help of renormalization group
equations~\cite{ross,nath,faraggi,castano,barger,kolda,sakis,pierce}.
The condition of REWSB 
puts an upper bound on $M_0$ of about $M_0 \lesim $ 5 TeV
(depending on the values of the other four parameters). 

In order to obtain a precise prediction for $\mh$ within the M-SUGRA
scenario, we employ the complete \twol\ RG running with appropriate
thresholds (both logarithmic and finite for the gauge couplings
and using the so called $\theta$-function approximation for 
the masses~\cite{sakis}) including full \onel\ minimization
conditions for the effective potential, in order to
extract all the parameters of the M-SUGRA scenario at the EW
scale. This method has been combined with the presently most precise
result of $\mh$ based on a Feynman-diagrammatic
calculation~\cite{mhiggsletter,mhiggslong}.
This has been carried out by combining the codes of two
programs namely, 
{\tt SUITY}~\cite{sakis2} and {\tt FeynHiggs}~\cite{feynhiggs}.

%%%%%%%%%%%%%%%%%%
\begin{figure}
\centerline{\psfig{figure=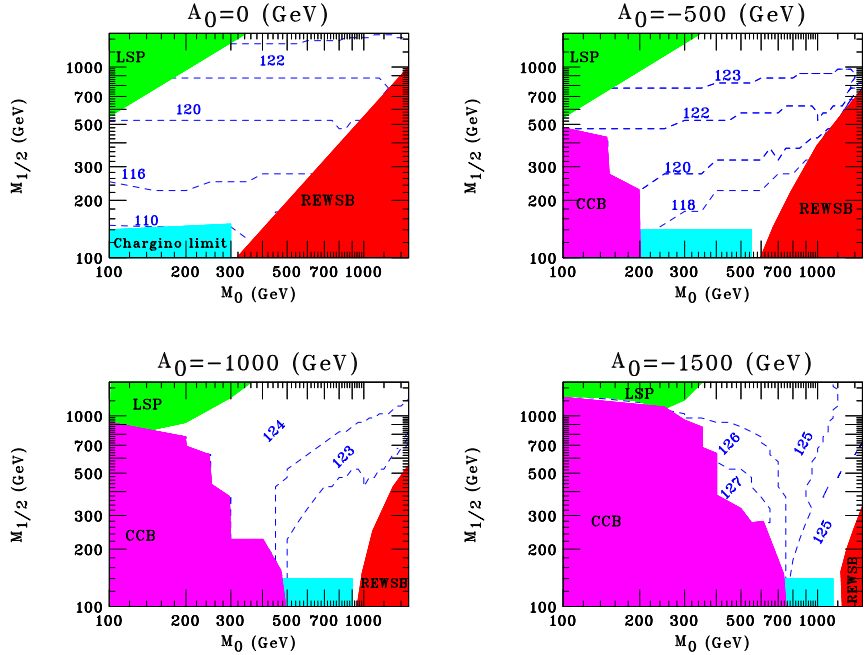,height=4.5in,angle=90}}
\caption{
In the $M_0-M_{1/2}$-plane the contour lines of $\mh$ are shown
for four values of $A_0$. The numbers refer to $\mh$ in the
respective region within $\pm 0.5 \gev$. The regions that are excluded
by REWSB, the CCB or LSP conditions, or by direct chargino search are
also indicated.
}
\label{fig:msugra}
\end{figure}
%%%%%%%%%%%%%%%

In order to investigate the upper limit on the Higgs boson mass in the
M-SUGRA scenario, we keep $\tb$ fixed at a large value, $\tb = 30$.
Concerning the sign of the Higgs mixing parameter, $\mu$, we find larger
$\mh$ values (compatible with the constraints discussed below) for 
negative $\mu$ (in the convention of \refeq{eq:xt}).
In the following we analyze the upper limit on $\mh$ as a function of 
the other M-SUGRA parameters, $M_0$, $M_{1/2}$ and $A_0$. 
Our results are displayed in \reffi{fig:msugra} for four values of
$A_0$: $A_0 = 0, -500, -1000, -1500 \gev$. We show contour lines of
$\mh$ in the $M_0-M_{1/2}$-plane. The numbers inside the plots indicate the
lightest Higgs boson mass in the respective area within $\pm 0.5 \gev$.
The upper bound on the lightest $\cp$-even Higgs boson mass 
is found to be at most 127 GeV. This upper limit is reached for 
$M_0 \approx 500 \gev$, $M_{1/2} \approx 400 \gev$ and $A_0 = -1500 \gev$.
Concerning the analysis the following should be noted:
\begin{itemize}
\item 
We have chosen the current experimental central value for
the top quark mass, $\mt = 174.3$ GeV. As mentioned above, increasing 
$\mt$ by 1 GeV results in an increase of $\mh$ of approximately $1 \gev$.
\item 
The M-SUGRA parameters are taken to be real, no SUSY $\cp$-violating
phases are assumed.
\item  
We have chosen negative values for the trilinear 
coupling, because $\mh$ turns out to be increased
by going from positive to negative values of $A_0$. 
$|A_0|$ is restricted from
above by the condition that no negative squares of squark masses and no
charge or color breaking minima appear.
\item
The regions in the $M_0-M_{1/2}$-plane that are excluded for the
following reasons are also indicated:
\begin{itemize}
\item
REWSB: parameter sets that do not fulfill the REWSB condition.
\item
CCB: regions where charge or color breaking minima occur
or negative squared squark masses are obtained at the EW scale.
\item
LSP: sets where the lightest neutralino is not the LSP. Mostly there the
lightest scalar tau becomes the LSP.
\item
Chargino limit: parameter sets which correspond to a chargino mass
that is already excluded by direct searches.
\end{itemize}
\item 
We do not take into account the $b\rightarrow s\gamma$
constraint as the authors of~\citere{pierce2,deboer} do. 
This could reduce the upper limit but still
the experimental and theoretical uncertainties of this constraint
are quite large.
\end{itemize}

%%%%%%%%%%%%%%%%%%%%%%%%%%%%%%%%%%%%%%%%%%%%%%%%%%%%%%%%%%%%%%%%%%%
%%%%%%%%%%%%%%%%%%%%%%%%%%%%%%%%%%%%%%%%%%%%%%%%%%%%%%%%%%%%%%%%%%%

\section{Conclusions}

We have analyzed the upper bound on $\mh$ within the MSSM. Using the
Feynman-diagrammatic result for $\mh$, which contains new
genuine two-loop corrections, leads to an increase of $\mh$ of up to 
$4 \gev$ compared to the previous result obtained by renormalization
group methods. We have furthermore investigated the MSSM parameters for
which the maximal $\mh$ values are obtained and have compared the 
$\mhmax$ scenario with the previous benchmark scenario. For $\mt =
174.3$~GeV and $\msusy = 1$~TeV we find $\mh \lesim 129$~GeV as upper
bound in the MSSM.
In case that no evidence of a Higgs signal is found before the
end of running in 2000, experimental searches for the Higgs boson at
LEP can ultimately be reasonably expected to exclude $\mh\lesim
113~\Gcs$. In the context of the $\mhmax$ benchmark scenario
(with $\mt = 174.3$~GeV, $\msusy = 1$~TeV) this rules out the interval 
$0.5\lesim\tb\lesim 2.4$ at the 95\% confidence level within the MSSM.
Within the M-SUGRA scenario, the upper bound on $\mh$ is
found to be $\mh \lesim 127 \gev$ for $\mt = 174.3 \gev$. 
This upper limit is reached for the M-SUGRA parameters $M_0 \approx 500
\gev$, $M_{1/2} \approx 400 \gev$ and $A_0 = -1500 \gev$. The upper
bound within the M-SUGRA scenario 
is lower by 2 and 4~GeV than the bound obtained in the general
MSSM for $\msusy = 1 \tev$ and $\msusy = 2 \tev$, respectively.

\section*{Acknowledgements}
A.D.\ acknowledges financial support from the Marie Curie Research
Training Grant ERB-FMBI-CT98-3438. A.D.\ would also like to thank Ben
Allanach for useful discussions. P.T.D.\ would like to thank Jennifer
Kile for providing the Standard Model Higgs boson production
cross-sections. G.W.\ thanks C.E.M.~Wagner for useful discussions.

%%%%%%%%%%%%%%%%%%%%%%%%%%%%%%%%%%%%%%%%%%%%%%%%%%%%%%%%%%%%%%%%%%%%%%%%%
%%%%%%%%%%%%%%%%%%%%%%%%%%%%%%%%%%%%%%%%%%%%%%%%%%%%%%%%%%%%%%%%%%%%%%%%%

\end{document}